\documentclass[aps,prl,twocolumn,twoside,amsmath,amssymb,bbm,superscriptaddress]{revtex4-1}
\usepackage[english]{babel}
\usepackage{graphicx}

\DeclareMathOperator{\tr}{tr}
\newcommand{\la}{\langle}
\newcommand{\ra}{\rangle}
\newcommand{\be}{\begin{equation}}
\newcommand{\ee}{\end{equation}}

\begin{document}

\title{Quantum fluctuation theorems beyond two-point measurements}
\author{Kaonan Micadei}
\affiliation{Institute for Theoretical Physics I, University of Stuttgart, D-70550 Stuttgart, Germany}
\author{Gabriel T. Landi}
\affiliation{Instituto de F\'isica da Universidade de S\~ao Paulo,  05314-970 S\~ao Paulo, Brazil}
\author{Eric Lutz}
\affiliation{Institute for Theoretical Physics I, University of Stuttgart, D-70550 Stuttgart, Germany}

\begin{abstract}
We derive detailed and integral quantum fluctuation theorems for heat exchange in a quantum correlated bipartite thermal system using  the framework of dynamic Bayesian networks. Contrary to the usual two-projective-measurement scheme that is known to destroy quantum features, these fluctuation relations fully capture  quantum correlations and quantum coherence at arbitrary times. 
\end{abstract}
\maketitle{}

%%%%%%%%%%%%%%%%%%%%%%%%%%%%%%%%%%%%%%
% INTRODUCTION
%%%%%%%%%%%%%%%%%%%%%%%%%%%%%%%%%%%%%%
Fluctuation theorems are fundamental generalizations of the second law of thermodynamics for small systems. While the entropy production $\Sigma$ is a nonnegative deterministic quantity for macroscopic systems, it becomes random at the microscopic scale owing to the presence of nonnegligible thermal \cite{sek10,sei12} or quantum  \cite{esp09,cam11} fluctuations. Detailed fluctuation theorems quantify the probability of occurrence of negative entropy production events via the general relation $P(\Sigma)/P(-\Sigma) = \exp(\Sigma)$ \cite{eva02}. Integral fluctuation theorems take on  the form $\la \exp(-\Sigma)\ra =1$ after integration over $\Sigma$. The concavity of the exponential function then implies that the entropy production is only positive on average, $\la \Sigma \ra \geq 0$. The generic validity of fluctuation theorems arbitrarily far from equilibrium makes them particularly useful in nonequilibrium physics. They have been extensively investigated for this reason, both theoretically and experimentally, for classical systems \cite{jar11,cil13}. These studies have provided unique insight into the thermodynamics of microscopic systems,  from colloidal particles to enzymes and molecular motors \cite{sek10,sei12}.

The situation is more involved in the quantum regime. Quantum fluctuation theorems are commonly studied within the  two-point-measurement (TPM) scheme \cite{esp09,cam11}. In this approach, the energy change, and in turn the entropy production, of a quantum system are determined for individual realizations   by projectively measuring the energy at the beginning and at the end of a nonequilibrium protocol \cite{tal07}. 
Equivalent formulations based on Ramsey-like interferometry  \cite{maz13,dor13} and generalized measurements \cite{ron14} have also been proposed. These  methods were used to perform 
experimental tests of quantum fluctuations theorems, both for mechanically  driven \cite{bat14,an15,cer17} and thermally driven \cite{pal18} systems, using NMR, trapped-ion and cold-atom setups. 
The TPM procedure successfully captures the discrete quantum  energy spectrum of the system, as well as its nonequilibrium quantum dynamics between the two measurements \cite{jar15}. However,  due to its projective nature, it completely fails to account for quantum correlations and quantum coherence, two central  features of quantum theory, that may be present in initial and final states of the system. In that sense, the TPM scheme  may thus be viewed as not fully quantum.  

In this paper, we present detailed and integral quantum fluctuation theorems for heat exchange  between  quantum correlated bipartite thermal systems using a dynamic Bayesian network approach \cite{nea03,dar09}. Global and local descriptions of a composite system usually differ because  of quantum correlations. The dynamic Bayesian network offers a powerful framework to specify the local dynamics conditioned on the global states, hence preserving all the quantum properties of the system, including quantum correlations and quantum coherence, in contrast to the TPM strategy. Our findings reduce to the Jarzynski-W\'ojcik fluctuation theorem in the absence of correlations \cite{jar04} and to the exchange fluctuation theorem of Jevtic and coworkers in the presence of classical correlations \cite{jev15}. They additionally complement recent attempts to obtain fully quantum fluctuation theorems for mechanically driven systems \cite{alh16,abe18,par17,Santos19} (see also Refs.~\cite{man18,kwo19}). 

In the following, we first derive a detailed quantum fluctuation theorem for the ratio of the probability of a conditional local  trajectory of the system and its reverse. We show that it accounts both for quantum correlations (in the form of a stochastic quantum mutual information \cite{nie00}) and for quantum coherence (in the form of a stochastic relative entropy of coherence \cite{bau14}). We further identify a contribution to the  entropy production that stems from the randomness of the conditional local trajectory. 
Moreover,  we obtain a detailed fluctuation relation for the joint probability of all  quantum contributions and demonstrate that each of them, as well as their sum, individually satisfies an  integral fluctuation theorem. 
Finally, we derive a modified quantum fluctuation relation for the heat variable alone, valid for any intermediate times.

%%%%%%%%%%%%%%%%%%%%%%%%%%%%%%

\textit{Dynamic Bayesian networks.} We consider  two arbitrary quantum systems, $A$ and $B$, with respective Hamiltonians $H_A$ and $H_B $,
 initially prepared in the joint state,
\begin{equation}\label{initial_state}
\rho_{AB}(0) = \rho_{A}^\text{0} \otimes \rho_B^\text{0} + \chi_{AB},
\end{equation}
where $\rho_i^\text{0} = \exp({-\beta_i H_i})/Z_i$ $(i= A,B)$ are local thermal Gibbs states at  inverse temperatures $\beta_i$ and $Z_i = \tr[\exp(-\beta_i H_i)]$ is the corresponding  partition function (see Fig.~1). 
The operator $\chi_{AB}$ induces correlations between the two subsystems. It is assumed to satisfy $\tr_i[\chi_{AB}]=0$, so that the reduced states, $\rho_i(0) = \tr_j[\rho_{AB}(0)]$, are locally thermal  even though $A$ and $B$ are globally correlated \cite{jev15}. This condition guarantees that the local systems have a well defined temperature.
Thermal contact between the two  systems is established at $t=0$ by letting them interact via an energy conserving unitary transformation $U(t)$ verifying $[U(t), H_A + H_B] = 0$. The global basis $|s_n\rangle$ at time $t_n$ is defined as $\rho_{AB}(t_n) = U(t_n) \rho_{AB}(0) U^\dagger(t_n) = \sum_s P_s |s_n\rangle\langle s_n|$, where $P_s$ is the initial population. On the other hand, the corresponding local bases, $|a_n\rangle$ and $|b_n\rangle$, follow from the decomposition of the reduced states, $\rho_A(t_n) = \sum_{a_n} P_{a_n} |a_n\rangle\langle a_n|$ and $\rho_B(t_n) = \sum_{b_n} P_{b_n} |b_n\rangle\langle b_n|$. We note that while the evolution of the global state is deterministic, with each eigenstate $|s\rangle$ simply  evolving in time according to $ U(t) |s\rangle$ and the initial populations $P_s$ kept fixed, that of the local (reduced) states is stochastic.

Our aim is to assess the statistics  of the heat exchanged between $A$ and $B$ at any given time, accounting for all the quantum properties of the process, including quantum correlations and  quantum coherence. This endeavor faces a number of mathematical and physical difficulties. Mathematically, the global state is not diagonal in the energy representation because of the nonvanishing correlations. As a result, the global and local bases are not mutually orthogonal, $\la a_nb_n|s_n\ra\neq \delta_{a_nb_n,s_n}$, making their relationship nontrivial, except when $\chi_{AB}=0$. The physical consequence is that the local bases, in which the exchanged heat variable is  evaluated, do not contain the complete information about the composite system.

%%%%%%%%%%%%%%%%%%%%%%%%%%%%%%

In order to solve these issues, we employ the tools of dynamic Bayesian networks which are widely used in computer science and statistics \cite{nea03,dar09}. They can be regarded as generalizations of hidden Markov models \cite{str60} which have been used to study classical fluctuation relations in the presence of hidden degrees of freedom \cite{kaw13,ehr17} (see also Refs.~\cite{ito13,str19}). These techniques allow the systematic analysis of  probabilities of  events conditioned on some other events. Concretely, at any given time $t_n$, the conditional probability  of finding the local systems $A$ and $B$ in their respective energy eigenstates $|a_n\rangle$ and $|b_n\rangle$, given that the global system is in state $|s_n\rangle$, is,
\begin{equation}\label{augmented_conditional}
P(a_n,b_n | s_n) = |\langle a_n\,b_n| s_n\rangle|^2.
\end{equation}
For any  sequence of  times,  $t_1, t_2, \ldots t_N$, we may define a conditional trajectory ${\Gamma} = (s,a_0,b_0, a_1,b_1,\ldots, a_N,b_N)$ (see Fig.~2) and the corresponding path probability as, 
\begin{equation}\label{path_prob}
\mathcal{P}[\Gamma] = P_s P(a_0,b_0 |s) P(a_1,b_1|s_1) \ldots P(a_N,b_N | s_N).
\end{equation}
The corresponding  probability for the  local trajectory is obtained by summing over all quantum trajectories $s$,
\begin{equation}
\mathcal{P}(a_0,b_0,\ldots,a_N,b_N) = \sum\limits_s \mathcal{P}[\Gamma].
\end{equation}
We may analogously introduce a reversed conditional local trajectory $\Gamma^*=(s^*,a_N,b_N,\ldots,a_0,b_0)$  with path probability $\mathcal{P}[\Gamma^*] =  P_{s^*} \bar P(a_N,b_N | s^*) \ldots  \bar P(a_0,b_0 |s_N^*)$ where
$\bar P (a_n,b_n | s_{N-n}^*) = |\langle a_n\,b_n | U^\dagger (t_{N-n}) | s^* \rangle|^2$.

\begin{figure}[t]
\centering
\includegraphics[width=0.43\textwidth]{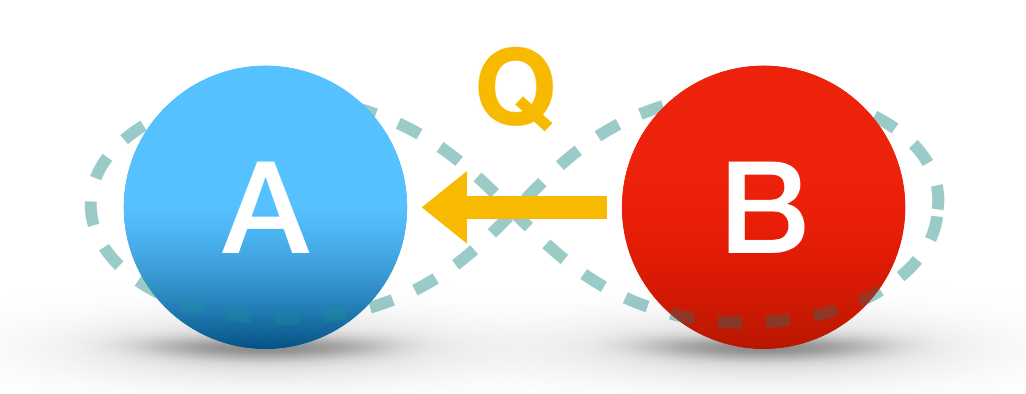}
\caption{\label{fig:schema}
Quantum correlated bipartite quantum system $AB$ in local thermal states at different temperatures. The initial joint state is of the form $\rho_{AB}(0) = \rho_{A}^\text{0} \otimes \rho_B^\text{0} + \chi_{AB}$ with Gibbs states, $\rho_i^\text{0} = \exp({-\beta_i H_i})/Z_i$ at inverse temperatures $\beta_i$ $(i= A,B)$, and initial quantum correlations $\chi_{AB}$. During thermal interaction, the two arbitrary subsystems exchange the amount of stochastic heat $Q$.
}
\end{figure}

For concreteness and simplicity, we shall next focus  on the case of a two-time probability, taken to be the initial time $t=0$ and an arbitrary future time $t_1$. Generalizations to multiple times are straightforward.
Marginalizing the conditional probability (\ref{path_prob}) over $a_0,b_0$ then yields, 
\begin{equation}
\mathcal{P}(a_1,b_1) = \sum\limits_{s,a_0,b_0} \mathcal{P}[\Gamma] = \langle a_1, b_1 | \rho_{AB}(t_1) | a_1,b_1\rangle, 
\end{equation}
which is  the result one would have expected on physical grounds. We furthermore  have the two probabilities ${\small \mathcal{P}}(a_1) = \sum_{b_1} \langle a_1, b_1 | \rho_{AB}(t_1) | a_1,b_1\rangle$ and ${\small \mathcal{P}}(b_1) = \sum_{a_1} \langle a_1, b_1 | \rho_{AB}(t_1) | a_1,b_1\rangle$.
Similarly, by only marginalizing over the global trajectory $s$, we obtain the path probability for the  local trajectory $(a_0,b_0,a_1,b_1)$,
\begin{equation}\label{latent_path_prob}
\mathcal{P}(a_0,b_0,a_1,b_1) = \sum\limits_s P_s P(a_0,b_0|s) P(a_1,b_1|s_1).
\end{equation}
Interestingly, these probabilities may also be cast in terms of the expectation value of a Choi matrix \cite{sup}. 
In the particular case where the initial state~(\ref{initial_state}) is separable ($\chi_{AB}=0$),  global and local bases are identical, $|s\ra=|ab\ra$, and
Eq.~(\ref{latent_path_prob})  reduces to the TPM result \cite{jar04}, 
\begin{equation}
\label{latent_path_prob_diag}
\mathcal{P}(a_0,b_0,a_1,b_1) = P_a^0 P_b^0 |\langle a_1,b_1 | U(t) | a_0,b_0\rangle|^2.
\end{equation}
Expression \eqref{latent_path_prob} hence generally contains  more information about the local quantum dynamics  than Eq.~\eqref{latent_path_prob_diag}.

%%%%%%%%%%%%%%%%%%%%%%%%%%%%%%

\begin{figure}
\centering
\includegraphics[width=0.49\textwidth]{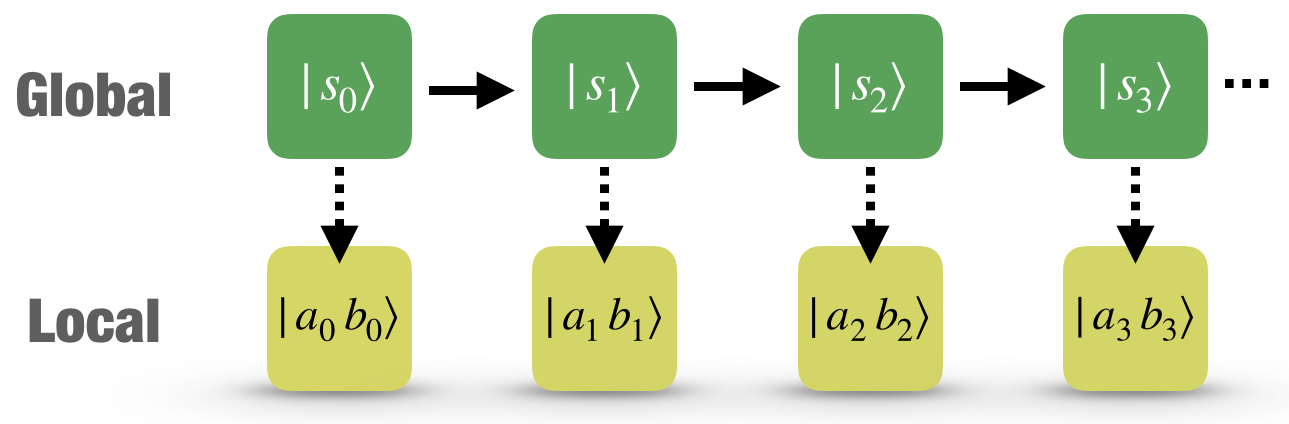}
\caption{\label{fig:latent_trajs}
Dynamic Bayesian network. 
The global quantum trajectory is specified by the  state $|s(t)\rangle$ which evolves deterministically.  At each instant  $t_n$,
the conditional probability of finding the reduced systems, $A$ and $B$, in their local energy eigenstates $|a_n,b_n\rangle$, given the state $|s_n\ra$, is specified by Eq.~(\ref{augmented_conditional}).
The set of points $(s,a_0,b_0, a_1,b_1,\ldots)$ defines a conditional local  trajectory $\Gamma$, with path probability ${\cal P}[\Gamma]$, Eq.~\eqref{path_prob}, that accounts for the  full quantum properties of the system.
}
\end{figure}
\textit{Detailed quantum fluctuation theorem.} We next derive a detailed fluctuation theorem for the ratio of  forward and reversed conditional trajectories using Eq.~\eqref{path_prob},
\begin{equation}
\label{8}
  \frac{{\cal P}[\Gamma]}{{\cal P}[\Gamma^*]}
  = \frac{P_s}{P_{s^*}} 
  \frac{P(a_0,b_0|s)P(a_1,b_1|s_1)}{\bar P(a_1,b_1|s^*) \bar P(a_0,b_0|s_1^*)}.
\end{equation}
In order to obtain an explicit expression for the theorem, we begin by rewriting the first ratio in Eq.~\eqref{8} as,
\begin{equation}
	\frac{P_s}{P_{s^*}}
    =
    \frac{P_{a_0} P_{b_0}}{P_{a_1} P_{b_1}} \exp\left(\ln \frac{P_{s}}{P_{a_0} P_{b_0}} - \ln \frac{P_{s^*}}{P_{a_1} P_{b_1}}\right),
\end{equation}
where $P_{a_1}$ and $P_{b_1}$ are the thermal occupations at time $t_1$. 
This then leads to  the quantum fluctuation relation,
\begin{equation}
\label{10}
	 \frac{{\cal P}[\Gamma]}{{\cal P}[\Gamma^*]}
    =
    \exp \left( Q_A \Delta\beta + I_0 - I_1 - \Sigma_A - \Sigma_B + \gamma  \right).
\end{equation}
We have here identified (i) the entropy production associated with heat exchange, $Q_A \Delta\beta =  (E_{a_1}- E_{a_0})(\beta_A-\beta_B)$, where $E_{a_n}$ are the  eigenenergies of  $H_A$, (ii)  the stochastic quantum mutual information, $I_0=\ln [{P_{s}}/{P_{a_0} P_{b_0}}]$, that accounts for  initial correlations between subsystems $A$ and $B$, and (iii) the stochastic quantum mutual information, $I_1=\ln [{P_{s^*}}/{\small \mathcal{P}(a_1) \mathcal{P}(b_1)}]$, that characterizes quantum correlations at the final time. We have additionally introduced the stochastic quantum relative entropies, $\Sigma_A=\ln [{\small \mathcal{P}(a_1)/P_{a_1}}]$ and $\Sigma_B=\ln [{\small \mathcal{P}(b_1)/P_{b_1}}]$. Finally, we have  discerned a  contribution to the entropy production, $\gamma = \ln [P(a_0,b_0|s)P(a_1,b_1|s_1)/\bar P(a_1,b_1|s^*) \bar P(a_0,b_0|s_1^*)]$, that comes from the second ratio in Eq.~\eqref{8}. This term stems from the stochastic nature of the conditional dynamics, in analogy to the classical result of Ref.~\cite{sei05}. It vanishes on average,  since the global dynamics is unitary and no extra energy is exchanged with an external bath.

Equation \eqref{10} is our first main  result. It generalizes  quantum  fluctuation theorems for heat exchange beyond the standard TPM approach \cite{jar04,jev15}. To make this point more precise, we express the stochastic quantum mutual informations, $I_l= J_l+ C_l$, $(l=0,1)$, as a sum of the stochastic classical mutual information, $J_l=\ln (P_{a_lb_l}/P_{a_l}P_{b_l})$, and of the stochastic quantum relative entropy of coherence, $C_l= \ln (P_s/P_{a_l b_l})$, which  is a proper measure of quantum coherence in a given basis \cite{bau14}.   The detailed fluctuation relation \eqref{10} therefore fully captures, at any time, the presence of quantum correlations between the two subsystems    and of quantum coherence, in the heat statistics. It provides, in particular, an extension of the fluctuation theorem of Jarzynski and W\'ojcik,  ${\small \cal P}[\Gamma]/{\small \cal P}[\Gamma^*] = \exp(Q_A \Delta \beta)$ \cite{jar04} and of Jevtic and coauthors,  ${\small \cal P}[\Gamma]/{\small \cal P}[\Gamma^*] = \exp(Q_A \Delta \beta - \Delta J)$ \cite{jev15}. 

By evaluating the average of the logarithm of Eq.~\eqref{10}, we furthermore obtain an expression for the mean heat exchanged between the subsystems $A$ and $B$,
\begin{equation}
\label{11}
\la Q_A \ra \Delta \beta= \Delta \la I\ra + S(\rho_A||\rho_A^0) + S(\rho_B||\rho_B^0),
\end{equation}
in agreement with the results of Ref.~\cite{mic19}. Equation \eqref{11} indicates that the heat current may be reversed, thus flowing from cold to hot, when the initial correlations are such that $\Delta \la I\ra + S(\rho_A||\rho_A^0) + S(\rho_B||\rho_B^0)\leq 0$. This process is enabled by a trade-off between correlations and entropy \cite{llo89}. The detailed fluctuation relation \eqref{10} extends this trade-off  to the level of individual quantum realizations.

\textit{Integral quantum fluctuation theorems.} An integral fluctuation relation that incorporates all the quantum contributions may be  derived from Eq.~\eqref{10} by integrating over all conditional trajectories $\Gamma$. We find,
\be
 \la \exp \left( Q_A \Delta\beta + I_0 - I_1 - \Sigma_A - \Sigma_B + \gamma  \right)\ra = 1.
 \ee
Interestingly, by using the  rules of Bayesian networks, one may show that each contribution satisfies an individual quantum fluctuation theorem \cite{sup}. We have, for example,
\begin{eqnarray}
		\!\!\!\!\!\!\langle e^{-I_0} \rangle
        &=& \sum_\Gamma {\cal P}[\Gamma] \exp\left({-\ln \frac{P_s}{P_{a_0}P_{b_0}}}\right)
        \\
               &=& \!\!\!\sum_{s,a_0,b_0} P(a_0,b_0|s)\,P_{a_0}P_{b_0}=\!\sum_{a_0,b_0} P_{a_0}P_{b_0} =1.
\end{eqnarray}
In a similar fashion (see Ref.~\cite{sup} for details), we obtain,
\be
\label{15}
\langle e^{-I_l} \rangle=\langle e^{-J_l} \rangle=\langle e^{-C_l} \rangle=\langle e^{-\Sigma_i} \rangle=\langle e^{-\gamma} \rangle=1.
\ee
We  therefore conclude that contributions from both classical and quantum correlations, $J_l$ and $I_l$,   as well as from quantum coherence, $C_l$,  separately obey an integral fluctuation relation, generalizing the recent findings of Refs.~\cite{ved12,xio18} for the quantum mutual information.  Equation \eqref{15} is our  second main result.

\textit{Modified detailed quantum fluctuation theorem for heat.} The detailed fluctuation relation \eqref{10} is formulated in terms of the probabilities of forward and reversed conditional trajectories. However, it is often convenient, both from a theoretical and an experimental point of view, to express it as a function of the joint probability of the different variables that appear in the exponent \cite{gar10,noh12,lah15}. To this end, it is important to separate variables according to their properties under time reversal \cite{lah15}. We therefore  introduce the odd (information) variable, $K=I_1 - I_0 + \Sigma_A + \Sigma_B $, and define the forward joint probability distribution of $K$, the odd variable $Q$ and $\gamma$ as $P_f(Q,K,\gamma) = \la \delta(Q-Q[\Gamma]) \delta(K-K[\Gamma,s^*])\delta(\gamma-\gamma[\Gamma,s^*])\ra$.  The corresponding reversed joint  probability distribution is  $P_r(-Q,-K,\bar\gamma) = \la \delta(Q-Q[\Gamma^*]) \delta(K-K[\Gamma^*,s])\delta(\gamma-\bar\gamma[\Gamma^*,s])\ra$ with  $\bar{\gamma}[\Gamma,s^*] = - \ln (|\langle a_0\,b_0 | s\rangle|^2\,|\langle a_1\,b_1 |U_t^\dagger|s\rangle|^2)/$ $(|\langle a_1\,b_1 | s^*\rangle|^2\,|\langle a_0\,b_0 |U_t |s^*\rangle|^2)$. The  relation \eqref{10} then implies the  detailed quantum fluctuation theorem \cite{sup},
\be
\label{16}
 \frac{P_f(Q,K,\gamma)}{P_r(-Q,-K,\bar\gamma)}
    =
    \exp \left( Q \Delta\beta - K + \gamma  \right).
\ee
\begin{figure}[ht]
\centering
\includegraphics[width=0.42\textwidth]{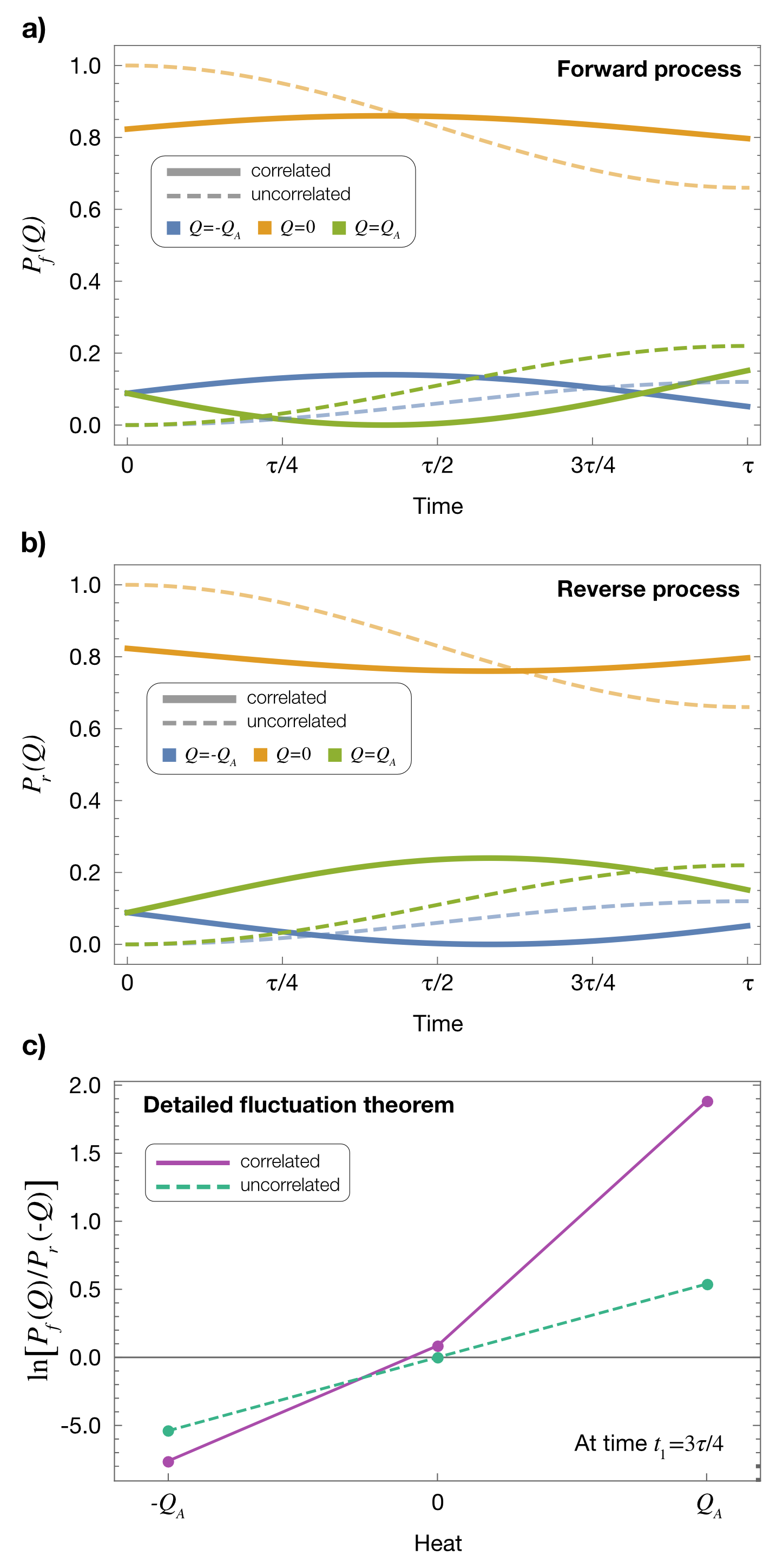}
\caption{\label{fig3}
Generalized quantum fluctuation theorem for heat for the two-spin-1/2 example. a) Forward quantum heat distribution $P_f(Q)$ for the three values $(0,\pm Q_A)$  with (thick lines) and without (thin lines) initial quantum correlations $\chi_{AB}$, as a function of the thermal interaction $\tau$.
b) Corresponding reversed heat distribution $P_r(Q)$. c) In the absence of initial correlations ($\alpha=0$), we have the Jarzynski-W\'ojcik   relation $P_f(Q)/P_r(-Q) = \exp(Q \Delta \beta)$ (green dashed line). On the other hand, in the presence of initial quantum correlations ($\alpha \neq 0$), we have the generalized fluctuation theorem,  $P_f(Q)/P_r(-Q) = \exp(Q \Delta \beta)/\Psi(Q)$ [Eq.~(17)] (purple solid line). The factor  $\Psi(Q)$ encapsulates the quantum features of the correlations and modifies the $Q$-dependence.}
\end{figure}
In like manner, a more general fluctuation relation of the form \eqref{16}  can be derived for all the individual quantum contributions by considering the joint probability distribution $ P_f(Q,J_0,C_0,J_1,C_1, \Sigma_A,\Sigma_B,\gamma)$. 
Integrating Eq.~(\ref{16}) over $K$ and $\gamma$, we eventually arrive  at the modified detailed quantum fluctuation relation for heat,
\be
\label{17}
 \frac{P_f(Q)}{P_r(-Q)} = \frac{\exp \left( Q \Delta\beta \right)}{\Psi(Q)},
\ee
where the factor $\Psi(Q) = \int d K d\gamma \; P(K,\gamma|Q) e^{-K - \gamma}$ depends on the correlations between $Q$, $K$ and $\gamma$. In the absence of correlations between the two subsystems $A$ and $B$, we recover the Jarzynski-W\'ojcik result, $\Psi_\text{JW}(Q)=1$ \cite{jar04}. The presence of quantum correlations thus modifies the exponential dependence on the heat variable on the right-hand side of Eq.~\eqref{17} through the function $\Psi(Q)$. This is our third main result.

\textit{Example.} Our findings are valid for arbitrary quantum systems. As an illustration, we now consider the case of an initially quantum correlated two-spin-1/2 system with Hamiltonians  $H_A = H_B = (1-\sigma_z)/2$, where $\sigma_z$ is the usual Pauli operator. This system has been recently investigated experimentally in a Nuclear Magnetic Resonance setup in Ref.~\cite{mic19}. The  correlation term in Eq.~\eqref{initial_state} is taken of the form $\chi_{AB} = \alpha\,|01\rangle\!\langle 10| + \alpha^*\,|10\rangle\!\langle 10|$ with parameter $\alpha$ \cite{mic19}. The value  $\alpha=0$ corresponds to initially uncorrelated local systems. We choose  $\alpha = -i \exp\left[-(\beta_A +\beta_B) /2 \right] \big/ (Z_A Z_B)$ for initial quantum correlations with nonzero geometric discord \cite{mic19}. We let the two subsystems interact, and exchange the amount of heat $Q$,  via the thermal operation $H_{int} = ({\pi}/{2\tau}) \left(\sigma_A^+ \sigma_B^- + \sigma_A^- \sigma_B^+ \right)$ for a time $\tau$. The thermal interaction induces four  transitions between the eigenstates of the two qubits, leading to three stochastic values of the heat, $Q = 0$ (twice) and $Q= \pm Q_A$, where $Q_A=  (E_{a_1}- E_{a_0})$ is the energy variation of spin $A$. 

We  analytically solve the respective global and local spin dynamics, and determine the forward and reversed heat distributions, $P_f(Q) = \sum_\Gamma \delta(Q-Q[\Gamma]) P[\Gamma]$ and $P_r(-Q) = \sum_{\Gamma^*} \delta(Q+Q[\Gamma^*]) P[\Gamma^*]$ \cite{sup}. The results are presented in Fig.~3  for $\exp({-\beta_A})/Z_A = 0.2$ and $\exp({-\beta_B})/Z_B = 0.3$. Figures 3ab show the forward and reversed quantum heat distributions for the three values $(0,\pm Q_A)$, with (thick lines) and without (thin lines) initial quantum correlations, as a function of the interaction $\tau$. We observe that the heat distributions depend explicitly on time and that the forward and reversed distributions are identical in the absence of initial correlations. Figure 3c displays the corresponding detailed quantum fluctuation relations for heat given by Eq.~\eqref{17}. Without initial correlations  ($\alpha =0$), we recover the Jarzynski-W\'ojcik fluctuation theorem which corresponds to $\Psi_\text{JW}(Q) =1$ (green dashed line). For $\alpha\neq 0$, the effect of the  quantum correlations  is clearly visible (purple solid line), modulating the $Q$-dependence via the function $\Psi(Q) \neq1$.

\textit{Conclusions.} We have used a dynamic Bayesian network approach to derive  detailed and integral  heat exchange fluctuation theorems for initially quantum correlated thermal bipartite systems. These fluctuation relations fully account for both quantum correlations and quantum coherence, two central quantum features, at arbitrary times, in contrast to the two-projective-measurement scheme. They provide much refined formulations of the second law of thermodynamics for small interacting quantum systems, compared to existing ones. We thus expect them to be useful for the study of far from equilibrium quantum thermodynamic systems.  

%%%%%%%%%%%%%%%%%%%%%%%%%%%%%

\textit{Acknowledgements}. We acknowledge financial support from  the S\~ao Paulo  Research Foundation  (Grants No. 2017/07973-5 and No. 2017/50304-7) and from the German Science Foundation (DFG) (Grant No. FOR 2724).

%%%%%%%%%%%%%%%%%%% SUPPLEMENTAL MATERIAL %%%%%%%%%%%%%%%%%%%%%%%%%%%%%
\clearpage
\widetext
\begin{center}
    \textbf{\large Supplemental Material: Quantum fluctuation theorems beyond two-point measurements}
\end{center}

\renewcommand\thesection{\Alph{section}}
\renewcommand\thesubsection{\arabic{subsection}}
\renewcommand{\theequation}{\thesection\arabic{equation}}

\setcounter{section}{0}
\setcounter{equation}{0}
\setcounter{figure}{0}
\setcounter{table}{0}
\setcounter{page}{1}
\makeatletter
\renewcommand{\theequation}{S\arabic{equation}}
\renewcommand{\thefigure}{S\arabic{figure}}
\renewcommand{\bibnumfmt}[1]{[S#1]}
\renewcommand{\citenumfont}[1]{S#1}

\section{A. Integral fluctuation theorems}

In this section, we present the derivations of   the individual integral fluctuation theorems given in Eq.~(15) of the main text. 
Special care should be paid  to the order with which sums are evaluated.

We first start with the final stochastic mutual information $I_1$. We have,
\begin{equation}
	\begin{split}
		\langle e^{-I_1} \rangle
        &= \sum_{\Gamma^*} P[\Gamma^*]\, \exp\left( -\ln \frac{P_{s^*}}{P_{a_1}\,P_{b_1}} \right)
        \\
        &= \sum_{s*,a_1,b_1}\, \sum_{a_0,b_0} |\langle a_1\,b_1|s^* \rangle|^2\,|\langle a_0\,b_0|U^\dagger(t)|s^* \rangle|^2 \,P_{a_1}\,P_{b_1}
        \\
        &= \sum_{a_1,b_1} \sum_{s^*} |\langle a_1\,b_1|s^* \rangle|^2\,P_{a_1}\,P_{b_1}
  = \sum_{a_1,b_1} P_{a_1}\,P_{b_1}
        = 1 .
	\end{split}
\end{equation}
Replacing the reversed path $\Gamma^*$  with the forward path $\Gamma$, a similar calculation shows that  the initial stochastic mutual information $I_0$ satisfies $\langle e^{-I_0} \rangle =1$. The classical component $J_1$ of the final stochastic mutual information  verifies,
\begin{equation}
	\begin{split}
		\langle e^{-J_1} \rangle
        &= \sum_{\Gamma^*} P[\Gamma^*]\, \exp \left(-\ln \frac{P(a_1,b_1)}{P_{a_1} P_{b_1}} \right)
        \\
        &= \sum_{s^*,a_1,b_1} \sum_{a_0,b_0} P_{s^*}\,|\langle a_1\,b_1|s^* \rangle|^2 \,|\langle a_0\,b_0|U^\dagger(t)|s^* \rangle|^2  \frac{P_{a_1} P_{b_1}}{P(a_1,b_1)}
        \\
        &= \sum_{a_1,b_1} \sum_{s^*} P_{s^*}\,|\langle a_1\,b_1|s^* \rangle|^2 \frac{P_{a_1} P_{b_1}}{P(a_1,b_1)}
        = \sum_{a_1,b_1} P_{a_1} \, P_{b_1}
        = 1 .
	\end{split}
\end{equation}
On the other hand, the calculation for the final stochastic relative entropy of coherence $C_1$ reads,
\begin{equation}
	\begin{split}
		\langle e^{-C_1} \rangle
        &= \sum_{\Gamma^*} P[\Gamma^*]\, \exp \left(-\ln \frac{P_{s^*}}{P(a_1,b_1)} \right)
        \\
        &= \sum_{s^*,a_1,b_1} \sum_{a_0,b_0} |\langle a_1\,b_1|s^* \rangle|^2 \,|\langle a_0\,b_0|U^\dagger(t)|s^* \rangle|^2  \, P(a_1 ,b_1)
        \\
        &= \sum_{a_1,b_1} \sum_{s^*} |\langle a_1\,b_1|s^* \rangle|^2 \, P(a_1,b_1)
        = \sum_{a_1,b_1} P(a_1,b_1)
        = 1 .
	\end{split}
\end{equation}
As before,  the integral fluctuation theorems for the initial stochastic classical mutual information $J_0$ and initial stochastic relative entropy coherence  $C_0$ follow by  taking the average over the forward path $\Gamma$.

We next turn to the local stochastic entropy productions, $\Sigma_A$ and $\Sigma_B$, during the forward  process $\Gamma$. We find, \begin{equation}
	\begin{split}
		\langle e^{-\Sigma_A} \rangle
        &= \sum_\Gamma P[\Gamma]\, \exp\left(-\ln \frac{\mathcal{P}_{a_1}}{P_{a_1}} \right)
        \\
        &= \sum_{s,a_1,b_1} \sum_{a_0,b_0} P_{s}\,|\langle a_0\,b_0|s \rangle|^2\,|\langle a_1\,b_1|U(t)|s \rangle|^2 \frac{P_{a_1}}{\mathcal{P}_{a_1}}
        \\
        &= \sum_{a_1,b_1} \sum_{s} P_{s}\,|\langle a_1\,b_1|U(t)|s \rangle|^2 \frac{P_{a_1}}{\mathcal{P}_{a_1}}
        =\sum_{a_1} \sum_{b_1} \mathcal{P}(a_1,b_1) \frac{P_{a_1}}{\mathcal{P}_{a_1}}
        = \sum_{a_1} P_{a_1}
        = 1 
	\end{split}
\end{equation}
and
\begin{equation}
	\begin{split}
		\langle e^{-\Sigma_B} \rangle
        &= \sum_\Gamma P[\Gamma]\, \exp\left(-\ln \frac{\mathcal{P}_{b_1}}{P_{b_1}} \right)
        \\
        &= \sum_{s,a_1,b_1} \sum_{a_0,b_0} P_{s}\,|\langle a_0\,b_0|s \rangle|^2\,|\langle a_1\,b_1|U(t)|s \rangle|^2 \frac{P_{b_1}}{\mathcal{P}_{b_1}}
        \\
        &= \sum_{a_1,b_1} \sum_{s} P_{s}\,|\langle a_1\,b_1|U(t)|s \rangle|^2 \frac{P_{b_1}}{\mathcal{P}_{b_1}}
        = \sum_{b_1} \sum_{a_1} \mathcal{P}(a_1,b_1) \frac{P_{b_1}}{\mathcal{P}_{b_1}}
        = \sum_{b_1} P_{b_1}
       = 1 .
	\end{split}
\end{equation}

Finally, the stochastic entropy production $\gamma$ satisfies an integral fluctuation theorem when averaging over the forward trajectory $\Gamma$,
\begin{equation}
	\begin{split}
		\langle e^{-\gamma} \rangle
        &= \sum_\Gamma P[\Gamma]\, \exp\left(-\ln \frac{|\langle a_0\,b_0|s \rangle|^2\,|\langle a_1\,b_1|U(t)|s \rangle|^2}{|\langle a_1\,b_1|s^* \rangle|^2\,|\langle a_0\,b_0|U^\dagger(t)|s^* \rangle|^2} \right)
        \\
        &= \left( \sum_{s} P_{s}\right) \left( \sum_{a_1,b_1} |\langle a_1\,b_1|s^* \rangle|^2 \right) \left( \sum_{a_0,b_0} |\langle a_0\,b_0|U^\dagger(t)|s^* \rangle|^2 \right)
        =1 .
	\end{split}
\end{equation}

\section{B. Detailed fluctuation theorem}

We next summarize the derivation of the detailed fluctuation theorem (16) of the main text. In order to evaluate the ratio $P_f(Q,K,\gamma)/P_r(-Q,-K,\bar\gamma)$, we need to consider that the forward trajectory $\Gamma$ is a function of $(s,a_0,b_0,a_1,b_1)$, while $Q[\Gamma]$ and $K$ and $\gamma$ are all functions of $(\Gamma,s^*)$. We first define,
\begin{equation}
    P_f(Q,K,\gamma | s^*) = \sum_\Gamma \delta(Q-Q[\Gamma])\, \delta(K-K[\Gamma,s^*])\,\delta(\gamma - \gamma[\Gamma,s^*])\, P(\Gamma)
\end{equation}
which gives the probability of having $(Q,K,\gamma)$ when one starts the reverse process with a vector $|s^*\rangle$. We have,
\begin{equation}
    P_f(Q,K,\gamma) = \sum_{s^*} P(s^*)\,P_f(Q,K,\gamma|s^*) .
\end{equation}
It then follows that,
\begin{equation}
 \begin{split}
 	P_f(Q,K,\gamma)
    &=
    \sum_{\Gamma,s^*} \delta(Q-Q[\Gamma])\, \delta(K-K[\Gamma,s^*])\,\delta(\gamma - \gamma[\Gamma,s^*])\, P(\Gamma)\,P(s^*)
    \\
    &=
    e^{Q\Delta\beta -K + \gamma} \sum_{\Gamma^*, s} \delta(Q+Q[\Gamma^*])\, \delta(K+K[\Gamma^*,s])\,\delta(\gamma - \bar\gamma[\Gamma^*,s])\, P(\Gamma^*)\,P(s)
    \\
    &=
    e^{Q\Delta\beta -K + \gamma} \sum_{s} P(s)\,P(-Q,-K,\bar \gamma | s)
    =
    e^{Q\Delta\beta -K + \gamma}\, P(-Q,-K,\bar \gamma) 
 \end{split}
 \end{equation}
 where
 $\bar{\gamma}[\Gamma,s^*] = - \ln \frac{|\langle a_0\,b_0 | s\rangle|^2\,|\langle a_1\,b_1 |U_t^\dagger|s\rangle|^2}{|\langle a_1\,b_1 | s^*\rangle|^2\,|\langle a_0\,b_0 |U_t |s^*\rangle|^2}$.
 
\section{C. Path probability for the local trajectory}
 
The physics behind expression (6) of the main text  for the path probability for the unconditional local trajectory 
 can be made more transparent by introducing a transformation akin to the Choi matrix used in the theory of quantum operations \cite{S_nie00}. 
We introduce an auxiliary Hilbert space $A'B'$ and consider
\begin{equation}\label{purification}
\Omega = \sum_s p_s \,|s\rangle\!\langle s|_{AB} \otimes |s\rangle\!\langle s|_{A'B'} .
\end{equation}
We then construct the Choi matrix, 
\begin{equation}
\Lambda(t) = (I_{AB} \otimes \mathcal{E}_{A'B'} ) (\Omega), 
\end{equation}
where $\mathcal{E}(\rho) = U(t) \,\rho\, U^\dagger(t)$. 
With simple rearrangements, Eq.~(6) of the main text may then  be written as,
\begin{equation}
\mathcal{P}(a,b,a',b') = \langle a,b,a',b' | \Lambda(t) | a,b,a',b'\rangle,
\end{equation}
which is in the form of a standard quantum mechanical expectation value. 
Since $\Lambda(t)$ is both Hermitian and positive semi-definite, the probabilities $\mathcal{P}(a,b,a',b')$ are guaranteed to be positive and normalized.

\section{D. Analytical solution of the  two-qubit example}
In this section, we provide the analytical solution for the two-spin example presented in the main text. For $\alpha=0$ the global initial state is
$\rho_{AB}(0) =  \mathrm{diag}\big(  1, e^{-\beta_B}, e^{-\beta_A}, e^{-\beta_A -\beta_B}\big)/({Z_A Z_B})$
where the diagonal is with respect to the $\sigma_z \otimes \sigma_z$ basis. From Eq.~(7) in the main text, the probability $P_f(Q)$  is given in this case by
$P_{\Gamma}(Q) = \sum_{\substack{a,a\\b,b'}} \delta(Q-\Delta E)\, \mathcal{P}(a,b) \,|\langle a',b' |U_t|a\,b\rangle|^2$. Under the action of the unitary $U_t = e^{-it H_{int}}$, the basis changes as follows,
\begin{align}
	U|00\rangle &= |00\rangle , \\
    U|01\rangle &= \cos\Big(t\frac{\pi}{2\tau}\Big)\,|01\rangle - i\sin\Big(t\frac{\pi}{2\tau}\Big) \,|10\rangle ,\\
    U|10\rangle &= - i\sin\Big(t\frac{\pi}{2\tau}\Big) \,|01\rangle + \cos\Big(t\frac{\pi}{2\tau}\Big)\,|10\rangle ,\\
    U|11\rangle &= |11\rangle .
\end{align}
Since initially the system $A$ is colder than system $B$, $Q=+Q_A$ when $|01\rangle \to |10\rangle$ and $Q=-Q_A$ when $|10\rangle \to |01\rangle$.  We have, therefore,
\begin{align*}
	P_f (Q=+Q_A) &= \mathcal{P}(0,1)\, |\langle 10 |U_t | 01 \rangle|^2 = \frac{e^{-\beta_B}}{Z_A Z_B} \sin^2\Big(t\frac{\pi}{2\tau}\Big),
    \\
    P_f (Q=-Q_A) &= \mathcal{P}(1,0)\, |\langle 01 |U_t | 10 \rangle|^2 = \frac{e^{-\beta_A}}{Z_A Z_B} \sin^2\Big(t\frac{\pi}{2\tau}\Big)  ,
    \\
    P_f (Q=0) &= \sum{a,b}\mathcal{P}(a,b)\, |\langle a\,b |U_t | a\,b \rangle|^2 = 
    		\frac{1 + e^{-\beta_A -\beta_B}}{Z_A Z_B} + \frac{e^{-\beta_A} 
            + e^{-\beta_B}}{Z_A Z_B} \cos^2\Big(t\frac{\pi}{2\tau}\Big)  .
\end{align*}
 For the reversed path $\Gamma^*$, the replacement $U_t \to U_t^\dagger$ implies the replacement $t \to -t$. In the uncorrelated case, this has no effect on the heat distribution and we have accordingly $P_f(Q) = P_r(Q)$.
 
On the other hand, in the correlated case when $\alpha = -i \exp\big[-(\beta_A + \beta_B)/2\big]/ Z_A Z_B$, the initial state reads,
 \begin{equation}
 	\rho_{AB}(0)  = 
    		\frac{1}{Z_A Z_B} |00\rangle\!\langle 00| 
            +  \frac{e^{-\beta_A} + e^{-\beta_B}}{Z_A Z_B} |\phi\rangle\!\langle \phi| 
            + \frac{e^{-\beta_A -\beta_B}}{Z_A Z_B} |11\rangle\!\langle 11|,
 \end{equation}
 with
 $|\phi\rangle = \left( e^{-\frac{\beta_B}{2}} |01\rangle + i\,e^{-\frac{\beta_A}{2}} |10\rangle \right) \big/ \sqrt{e^{-\beta_A } + e^{- \beta_B}}$.
 
 We have again,  $Q=+Q_A$ when $|01\rangle \to |10\rangle$ and $Q=-Q_A$ when $|10\rangle \to |01\rangle$. As a result,
\begin{align}
	P_f (Q=+Q_A) &= 
    		\mathcal{P}(\phi)\, |\langle 01 |\phi \rangle|^2 \,|\langle 10 |U_t | \phi \rangle|^2 
            = \frac{e^{-\beta_B}}{Z_A Z_B} 
            	\frac{\left[ e^{-\frac{\beta_A}{2}} \cos\Big(t\frac{\pi}{2\tau}\Big)  -  e^{-\frac{\beta_B}{2}}\sin\Big(t\frac{\pi}{2\tau}\Big) \right]^2}{e^{-\beta_A} + e^{-\beta_B}} ,
    \\
    P_f (Q=-Q_A) &= 
    		\mathcal{P}(\phi)\, |\langle 10 |\phi \rangle|^2 \,|\langle 01 |U_t | \phi \rangle|^2 
            = \frac{e^{-\beta_A}}{Z_A Z_B} 
            	\frac{\left[ e^{-\frac{\beta_B}{2}} \cos\Big(t\frac{\pi}{2\tau}\Big)  +  e^{-\frac{\beta_A}{2}}\sin\Big(t\frac{\pi}{2\tau}\Big) \right]^2}{e^{-\beta_A} + e^{-\beta_B}} ,
    \\
    P_f (Q=0) &= \mathcal{P}(0,0) + \mathcal{P}(1,1) + \mathcal{P}(\phi) 
        \Big( |\langle 01 |\phi \rangle|^2 \,|\langle 01 |U_t | \phi \rangle|^2
        + |\langle 10 |\phi \rangle|^2 \,|\langle 10 |U_t | \phi \rangle|^2 \Big) 
        \nonumber\\
        &= \frac{1 + e^{-\beta_A - \beta_B}}{Z_A Z_B}
        + \frac{e^{-\beta_A}}{Z_A Z_B} 
            	\frac{\left[ e^{-\frac{\beta_A}{2}} \cos\Big(t\frac{\pi}{2\tau}\Big)  -  e^{-\frac{\beta_B}{2}}\sin\Big(t\frac{\pi}{2\tau}\Big) \right]^2}{e^{-\beta_A} + e^{-\beta_B}} \nonumber\\
        &\qquad + \frac{e^{-\beta_B}}{Z_A Z_B} 
            	\frac{\left[ e^{-\frac{\beta_B}{2}} \cos\Big(t\frac{\pi}{2\tau}\Big)  +  e^{-\frac{\beta_A}{2}}\sin\Big(t\frac{\pi}{2\tau}\Big) \right]^2}{e^{-\beta_A} + e^{-\beta_B}} .
\end{align}
In general, except for $t = (0,\tau)$, $P_f(Q) \neq P_r(Q)$.

\end{document}